\documentclass[12pt,preprint]{aastex}
\usepackage{lscape}

\newcommand{\mpy}{mas yr$^{-1}$}

\begin{document}
\title{Investigation of the errors in SDSS proper-motion measurements
using samples of quasars}

\shorttitle{Proper-motion errors in SDSS}

\shortauthors{Dong et al.}

\author{Ruobing Dong\altaffilmark{1}, James Gunn\altaffilmark{1}, Gillian Knapp\altaffilmark{1}, Constance Rockosi\altaffilmark{2}, and Michael Blanton\altaffilmark{3}}

\altaffiltext{1}{Department of Astrophysical Sciences, Princeton University, Princeton, NJ 08544, rdong@astro.princeton.edu}

\altaffiltext{2}{UCO/Lick Observatory, Department of Astronomy and Astrophysics, University of California, Santa Cruz, CA 95064}

\altaffiltext{3}{Center for Cosmology and Particle Physics, Department of Physics, New York University, 4 Washington Place, New York, NY 10003}


\begin{abstract}

We investigate in detail the probability distribution 
function (pdf) of the 
proper-motion measurement errors in the SDSS+USNO-B
proper-motion catalog of \citet{mun04} using clean quasar samples. 
The pdf of the errors is well-represented by a Gaussian core with extended wings, plus a very small fraction ($<0.1\%$) of  ``outliers''. We find while formally the pdf could be well-fit by a five-parameter fitting function, for many purposes it is also adequately to represent the pdf with a one-parameter approximation to this function. We apply this pdf to the calculation
of the confidence intervals on the true proper motion for a 
SDSS+USNO-B proper motion measurement, 
and discuss several scientific applications 
of the SDSS proper motion catalogue. Our results have various applications in studies of the galactic structure and stellar kinematics. Specifically, they are crucial for searching hyper-velocity stars in the Galaxy.

\end{abstract}



\section{Introduction}\label{sec:introduction}

As of the eighth Data Release (hereinafter DRn, where n is the
release number) of the Sloan Digital Sky
Survey (hereinafter SDSS) 
the survey has released imaging for 14,555 deg$^2$,
or over a third of the sky \citep{gun98,yor00,lup01,sto02,aih11}. The imaging
catalog contains almost half a billion distinct detected objects down to
a $50\%$ completeness limit of $r=22.5$ for point sources,
and the survey's 1.8
million catalogued spectra provide several well defined samples of galaxies,
quasars, stars, and other objects \citep{aih11}.

Along with the photometric and spectroscopic data, SDSS also provides
proper-motion (PM) data \citep[hereinafter Munn et al.
catalog]{mun04}, produced by matching the
SDSS point source detections with earlier observations, including the
USNO reductions of the  Palomar Observatory
Sky Surveys (POSS-I and POSS-II), which span about 50 yr
in time \citep{monet03}. In this 
catalog, the USNO proper-motion system is re-calibrated
and made absolute
using SDSS galaxies, and these proper motions (called here SDSS+USNO-B PM) 
are computed
including both SDSS and USNO-B positions. The resultant catalog is $90\%$
complete to $g<19.7$, and  has a less than $0.5\%$ contamination rate.
The systematic errors are
on the order of 0.1 \mpy, and the statistical errors are roughly 3-4 \mpy ~in
each component of the PM. \cite{mun04} compared their results
with those of the revised New Luyten Two-Tenths catalog (rNLTT;
\citealt{gou03,sal03}). \cite{bon10} carried out a further comparison
of the SDSS+USBO-B proper motions with those of a sample of stars in the 
North Galactic Pole region \citep{maj92} and with proper motion  
measurements made using data from SDSS Stripe 82 \citep{bra08}.
These independent
measurements are expected to have different systematic errors from the
SDSS+USNO-B PM. The results show that the median differences and the rms
scatter of these comparisons agree with expectation
and that the SDSS+USNO-B PM
measurements are reliable at roughly the stated errors.

Proper-motion measurement at this level of accuracy are useful in several ways. 
Samples of objects can be defined using the reduced proper motion diagram
to separate classes of objects with intrinsically similar colors,
spectra and apparent magnitudes but very different proper motion 
distributions due, for example, to different luminosities or different
kinematics. This is used in
several target selection algorithms
in the SDSS projects, most notably the Sloan Extension for Galactic
Understanding and Evolution \citep[SEGUE I and II]{yan09}, both
to help find very nearby objects (with high PM) (e.g.
\citealt{lep08}) and distant giants
(with low PM). A variation of this method has been explored by defining a ``reduced-proper-motion'' (for example in the $r$-band the reduced-proper-motion would be defined as $r_{\rm RPM}=r+5\log{\rm PM}$, \citealt{sal03}), while the reduced proper motion diagrams are not as useful at faint magnitudes probed by SDSS as they are at the (traditionally used) bright end \citep{ses08}.
Most importantly, PM, along with even crude
distance determinations for stars in the Galaxy, provide two-dimensional
velocity information; if in addition radial velocity measurements
are available, one has full three-dimensional velocity information.
A one-sigma accuracy of 3 \mpy corresponds to a transverse velocity accuracy
of about 15 km/s at 1 kpc and 150 km/s at 10 kpc. Thus 
the PM measurements are
{\it statistically} very useful for studying the kinematics of the thick
disk and halo, for which the velocity dispersions are of this order at
these distances---provided one understands the errors well.
Recent years have witnessed an increased interest in
studies of stellar kinematics and Galactic structure, spurred both by
the increasing sophistication of galaxy formation simulations and
the availability of the large
photometric sample of stars in SDSS together with a significant subsample
with radial velocities and chemical composition information 
obtained as part of 
its SEGUE subsurveys. With the SDSS+USNO-B PM
measurement and radial velocities, \citet{bon10} selected a
sample of main-sequence stars with $r<20$, while  \citet{car10} selected a 
subsample of SDSS calibration stars (mostly metal-poor
turnoff F stars) to study the kinematics of both the galactic
thin and thick disks and the halo (see also \citealt{smi09,sch09,fuc09}). 
Many more studies of this sort are underway using the SEGUE II data.

To do this job really well clearly requires detailed knowledge of the
distribution of the proper motion, distance and radial velocity 
errors, since for most of these applications the
proper-motion related velocity errors are of the same order as the
velocities themselves. Furthermore, in most cases the error in the tangential velocity due to the proper motion and distance error is the dominant contributor to the total velocity error. In addition, if one is interested (and one
usually is) in extreme velocities, understanding the behavior 
of the distribution in the tails of the pdf is crucial. We
approach this problem in the way others have done in the past, but
with the goal from the outset of understanding the pdf in as much
detail as the sample sizes and systematics will allow.

Quasars are sufficiently distant that they should not have any measurable
PM in SDSS. Thus, the measured PM of quasars are just the PM measurement
error. By studying the PM of a large sample of quasars, we can probe the
statistical properties of the PM errors, and study the dependence on various
observational and instrumental parameters. Although the spectral 
energy distributions for quasars differ from those of the
point sources for which PM is of interest (stars), the major contributors
to the systematic error in the PM measurement, including: the difficulty in
centroiding sources on photographic plates; errors due to unresolved or
partially resolved projected nearby objects;
errors in the primary reference catalogs; systematic errors in the
UCAC catalog  \citep{zacharias10};  and charge transfer effects 
in the SDSS astrometric and photometric detectors
\citep{mun04}, work in essentially
the same way for all point sources. Thus the error distribution for all
point sources will be very similar (\citealt{bon10}, but caution is still
needed when applying the results based on quasars to stars, see
Section \ref{sec:fitting} and \ref{sec:summary} for details). Along with the publication of the
PM catalog, \citet{mun04} used spectroscopic quasars in the SDSS DR1
\citep{schneider02} to study the
mean and variance of the SDSS+USNO-B PM error, but due to the limited sample
size  were not able to study the dependence on magnitude. With a much
larger spectroscopic quasar sample in SDSS DR7, \citet{bon10} addressed this
issue again and presented the width $\sigma$ in the Gaussian distribution as
a function of $r$-band magnitude. They assumed the error distribution
was Gaussian and did not investigate the form of the distribution function;
in particular, they 
did not try to characterize the wings of the distribution. 

In this work, we analyze the
full error distribution of the SDSS+USNO-B PM (with the corrections
presented by
\citealt{mun08}, which corrected an error in the calculation of the proper
motions in right ascension). We do this using clean quasar samples which we
define in such a way that corresponding clean stellar samples are easily
constructed. We fit the PM error distribution
in the entire magnitude range by a Gaussian distribution for the core plus
a wing function, and derive the dependence of the function
parameters on magnitude.
Using this fitting function, we calculate the significance of a given
PM measurement. We also quantify (or at least place upper limits on) 
the fraction of the outliers in the PM measurement which survive the
cleaning process and for one reason or another clearly do not
belong to the main distribution. We then discuss various issues which involve
applying the analysis of the PM error distribution to other samples.

The structure of this paper is as follows. In Section~\ref{sec:sample} we
introduce the quasar samples and the criteria for selecting reliable
PM. We then fit the PM error distribution by parametrized analytic functions 
of both complex and and simplified forms in 
Section~\ref{sec:fitting}, and calculate
the significance of measured PM in Section~\ref{sec:significance}. We
summarize our results and discuss the issues related to using them in
Section~\ref{sec:summary}.



\section{Quasar Sample Selection}\label{sec:sample}

Our goal is to select a quasar sample with as few
contaminants as possible ({\it e.g.} stars and other objects with intrinsic
non-zero PM), while maintaining a sample size large enough for statistical
analysis, and at the same time covering the largest possible 
magnitude range to allow
studying the dependence of PM error on brightness. There are four quasar
samples in the recent literature which we have looked at: \citet{ric09} and
\citet{bov11}, which are photometric samples, and \citet{bon10} and
\citet{sch10}, which are spectroscopic samples. We expect that these samples
overlap with each other on most part (especially \citet{bon10} and
\citet{sch10}), but the non-overlaping part still makes significant difference
on the cleanness of the samples, as we will show. For each sample, we obtain the
PM and photometry from the SDSS DR7 Catalog Archive Server table {\tt propermotions}. 
We use the DR7 values in preference to the DR8 values both because of a known
error in the generation of the DR8 proper motions, and because
the photometry in the new DR8 reductions has been less well checked.
The DR8 PM problem is discussed
on the DR8 website and in \citet{aih11}, and will be repaired in DR9.
 
\begin{figure}[tb]
\begin{center}
\epsscale{0.4} \plotone{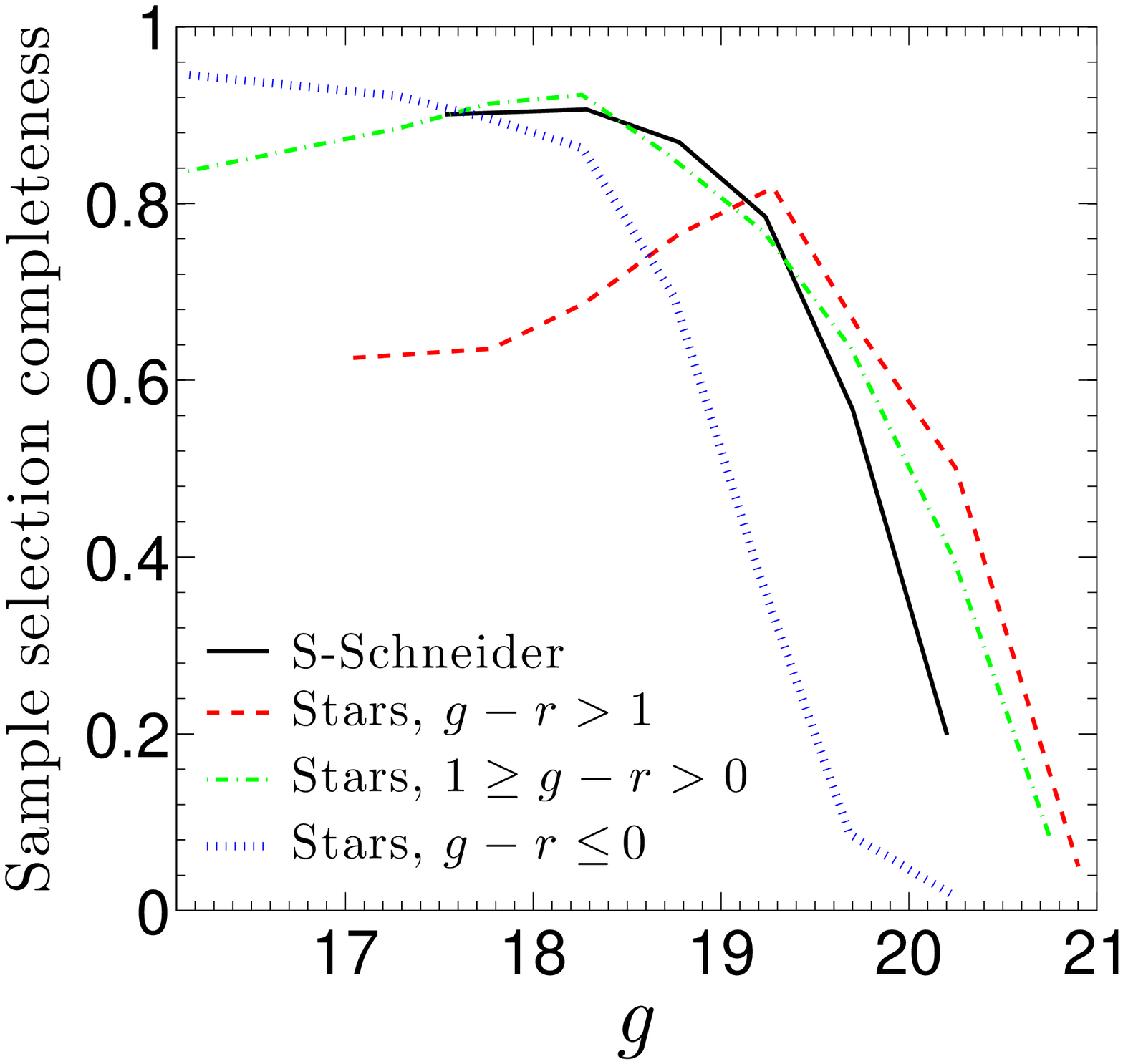}
\end{center} \figcaption{The clean PM sample selection completeness (defined as the ratio of the number of objects which survive our clean PM cut over the total number of objects) as a function of $g$-band magnitude for S-Schneider and all spectroscopic stars in SDSS DR7 (binned into three color bins).
\label{fig:completeness}}
\end{figure}

We use the following criteria \citep{kil06} to determine a {\it
clean} PM\footnote{See Catalog Archive Server for details of these
quantities http://cas.sdss.org/dr7/en/help/browser/browser.asp}:
\begin{enumerate}
\item {\tt match = 1}
\item {\tt sigRa < 525 \&\& sigDec < 525}
\item {\tt nFit = 6}
\item {\tt dist22 > 7}
\end{enumerate}
where {\tt match} is the number of objects in USNO-B which matches a SDSS
object within a 1 arcsec radius, {\tt sigRa} and {\tt sigDec} are the rms
residuals for the proper motion fit in right ascension and declination,
{\tt nFit} is the number of detections used in the fit including the SDSS
detection (so {\tt nFit = 6} requires that the object was detected on all five
USNO-B plates plus one for SDSS), and {\tt dist22} is the distance to the nearest
neighbor in SDSS with $g < 22$.
We rejected PM entries which violate any of the above conditions. Specifically, for our main quasar sample S-Schneider (see below), the fraction of objects which violates each cut condition is: $8.2\%$ for {\tt nFit=4} (the minimum {\tt nFit} in Munn et al. catalog), $17.3\%$ for {\tt nFit=5}, $0.5\%$ for {\tt match $\neq$ 1}, $2.9\%$ for {\tt sigRa $\geq$ 525} or {\tt sigDec $\geq$ 525} , and $5.6\%$ for {\tt dist22 $\leq$ 7}. The sample selection completeness (defined as the ratio of the number of objects which survive the cut over the total number of objects) as a function of the $g$-band magnitude (the six $g$ bins in Table \ref{tab:sample}) for S-Schneider is shown in Figure \ref{fig:completeness}. The selection completeness saturates at the bright end ($\sim90\%$), and drops rapidly with decreasing magnitude. The selection completeness curves for several samples of stars are shown in the same plot for comparison. We pick out all spectroscopic stars in SDSS DR7, bin them into three color bins ($g-r>1.0,1.0\geq g-r>0,g-r\leq0$), and carry out the same cleaning process to obtain the selection completeness as a function of $g$. As the color becomes redder the completenes for stars deceases at the bright end while increases at the faint end.

We note here that the way in which we trim the catalog for {\it clean} PM is
different from the widely used standard procedure ({\it e.g.} \citealt{bon10})
which is recommended in the original sample design \citep{mun04}, where it
only requires {\tt match = 1} and {\tt sigRa < 350 \&\& sigDec < 350}.
Considerations of various selection criteria
show that the standard PM selection criteria return a
much less clean quasar PM distribution, which has significantly more quasars
with spurious large PM values. Specifically, the important new condition {\tt nFit = 6} makes the cleanness of our samples much better than that of samples
defined using the standard criteria defined by \cite{mun04}, 
even when we loosen the requirement on rms fitting residuals
a little bit (see Table \ref{tab:sample} and discussion below).
We would like to stress that the 
functional forms which we derive later in this paper for the PM error
distribution
can only be applied to {\it clean} PM
samples as defined above.

We now use the several SDSS quasar catalogues from the literature to 
define samples for the analysis of the PM error distribution. 
\citet{ric09} selected $\sim1.2$ million photometric quasar candidates from
SDSS DR6 based on a Bayesian selection method employing the kernel density
estimate (KDE) of the probability density function. The $i$ magnitude range
for this sample is $\sim17-21$. After selecting all the objects which passed
their selection criterion, \citet{ric09} flagged the most likely
contaminants by assigning a {\tt good} index to every object. This index
starts at 0 for each object, then is incremented or decremented based on a set
of rules. In the end, every object is assigned
a {\tt good} index in the range of
[-6,6], with larger positive
values meaning a larger likelihood of being a quasar. For
our purpose, we back out the part of the {\tt good} index determination which
makes use of the PM, and assign each object a new {\tt good$^\prime$} index which
does not contain any PM information. We then choose objects with the new
{\tt good$^\prime$} $\geq3$ so that we obtain a clean sample 
(hereinafter S-Richards) while
retaining enough objects to carry out the analysis.

\citet{bov11} generated a SDSS quasar targeting catalog by assigning
a probability as a star, low-redshift ($z<2.2$), medium-redshift,
or high-redshift ($z>3.5$)
quasar for $\sim160$ million point sources with dereddened $i$-band magnitude
between 17.75 and 22.45 in SDSS DR8. They did this by modeling
the distributions
of stars and quasars in flux space down to the SDSS
flux limit by applying the
extreme-deconvolution method to estimate the underlying density
of each class as a function of magnitude, and then
convolved the densities with the flux uncertainties to assign
to each object the probability of its being a quasar.
We select a subsample (here after S-Bovy) with available clean
PM, and reject objects with {\tt good!=0} (objects fail on some
of the $BOSS$ flag cuts), {\tt Photometric=0} (objects were
observed under bad imaging condition) and quasar
probability $\leq99\%$ for all three categories.

\citet{bon10} selected 69,916 spectroscopic quasars from SDSS DR7 with
$14.5<r<20$ and $0.5<z<2.5$. We repeat the selection to define the sample
S-Bond  using the above improved criteria for
finding {\it clean} PM.

\citet{sch10} produced the spectroscopic quasar catalog for SDSS DR7, which
contains 105,783 spectroscopically confirmed quasars with luminosities larger
than $M_i=-22.0$. This catalog has been visually inspected, has highly
reliable redshifts from 0.065 to 5.46, and 
contains quasars fainter than $i\approx15$. We
select objects with clean PM from their catalog to form the sample
S-Schneider. Again, we expect a large overlapping between S-Schneider and S-bond.
Furthermore, to show the differences on the resulting samples due to different clean PM conditions, we select an additional sample S-Schneider-W from the Schneider quasar catalog with the recommended clean PM conditions in \citet{mun04} ({\tt match = 1} and {\tt sigRa < 350 \&\& sigDec < 350}).

Note that in all these samples, an object can have a large measured proper
motion for two reasons. First, it could in fact has a real large proper motion, and is 
therefore presumably not,
in fact, a quasar but a white dwarf or other peculiar star. These sneak
through even in the spectroscopic samples (as an example, the object
{\tt plate=1642, mjd=53115, fiber=81} is included in S-Schneider,
but actually it is a white dwarf). Second, the proper motion measurement has
occasionally failed for some reason, such as mismatches with USNO-B or bad deblends.
Clearly the assignment of a proper motion error
to the latter category or to the `tail' of the main distribution is 
a bit subjective, but in fact for our samples is pretty clear, as we 
shall see.

\begin{figure}[tb]
\begin{center}
\epsscale{0.49} \plotone{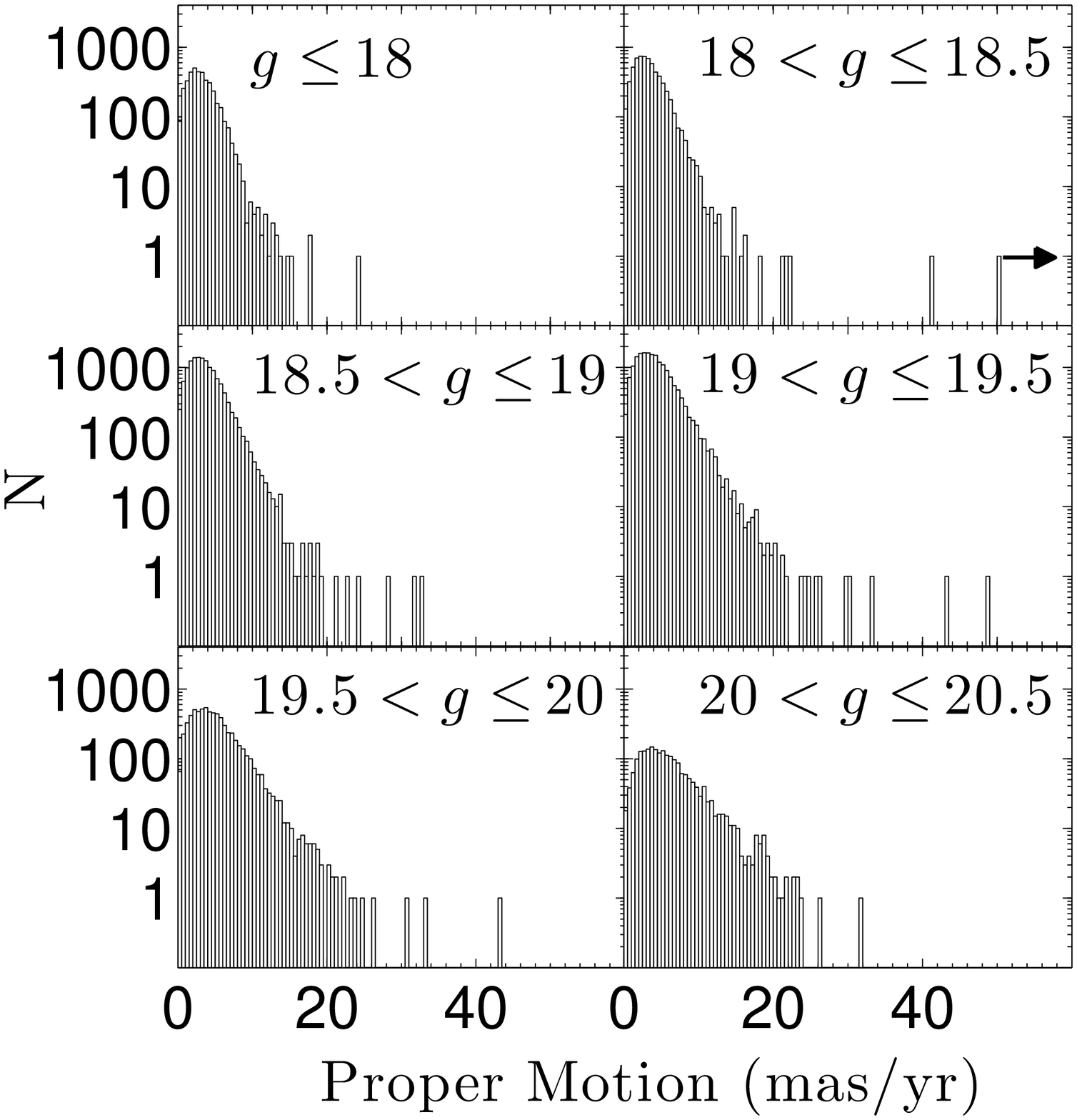} \plotone{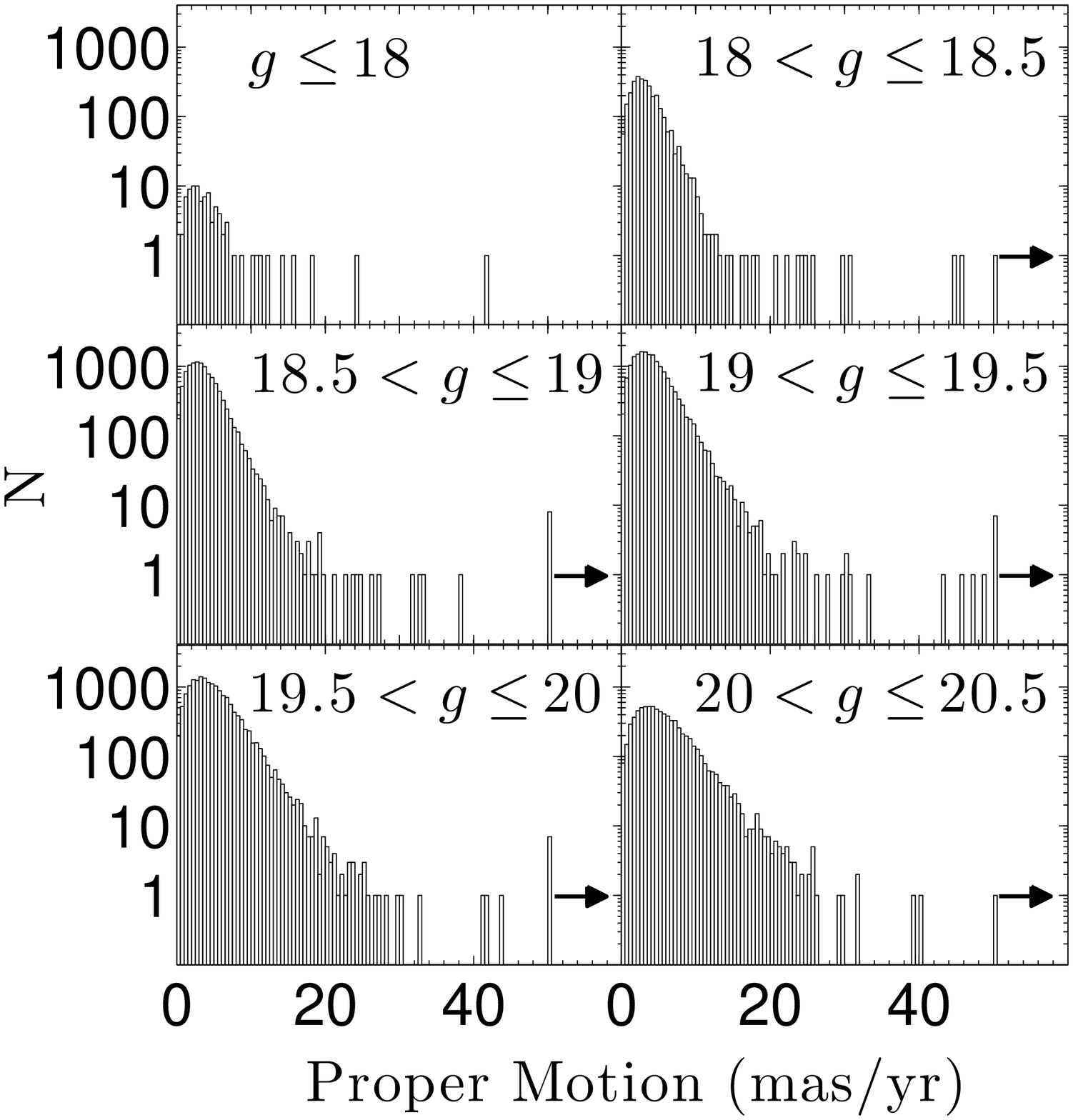} \plotone{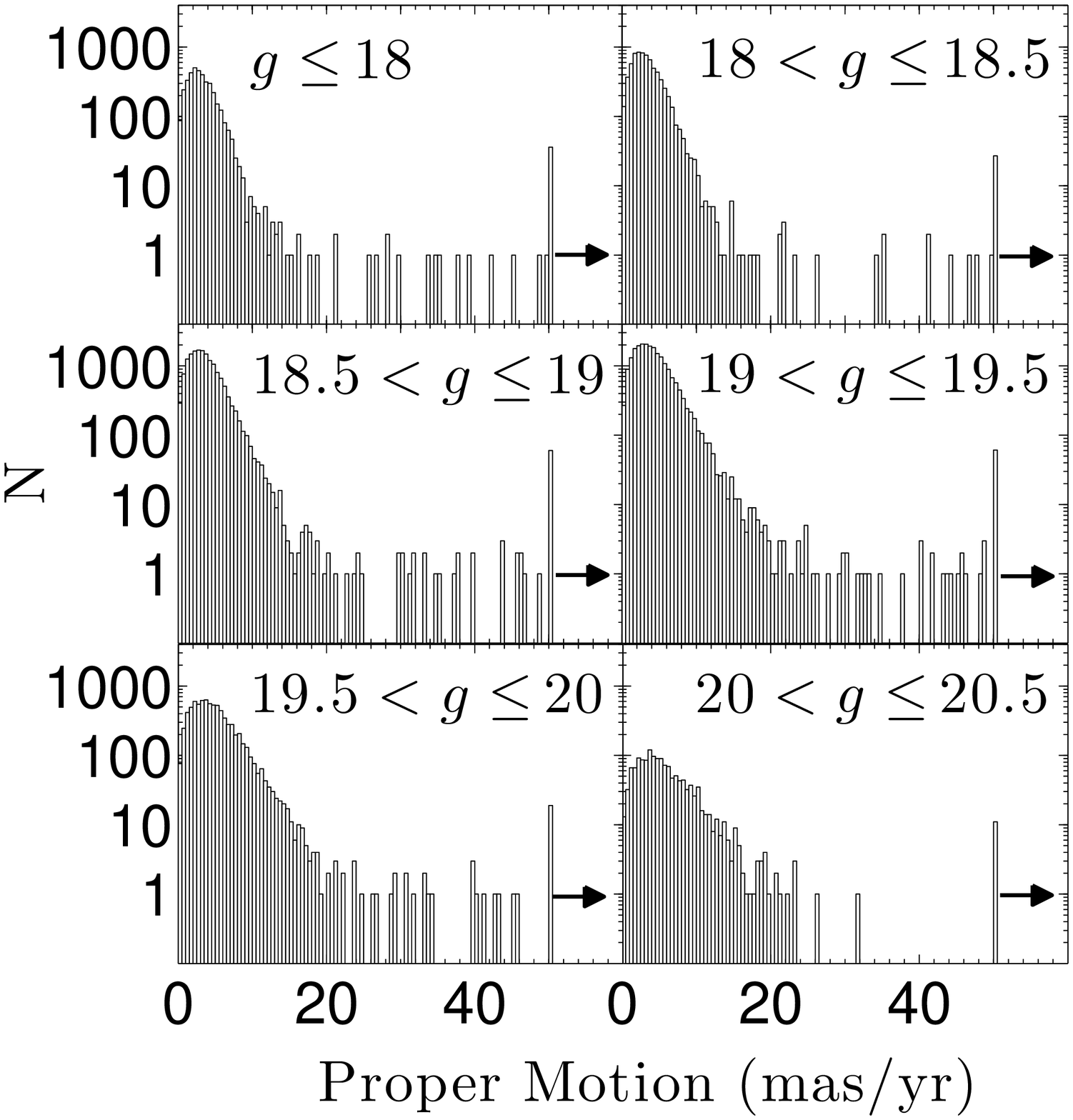} \plotone{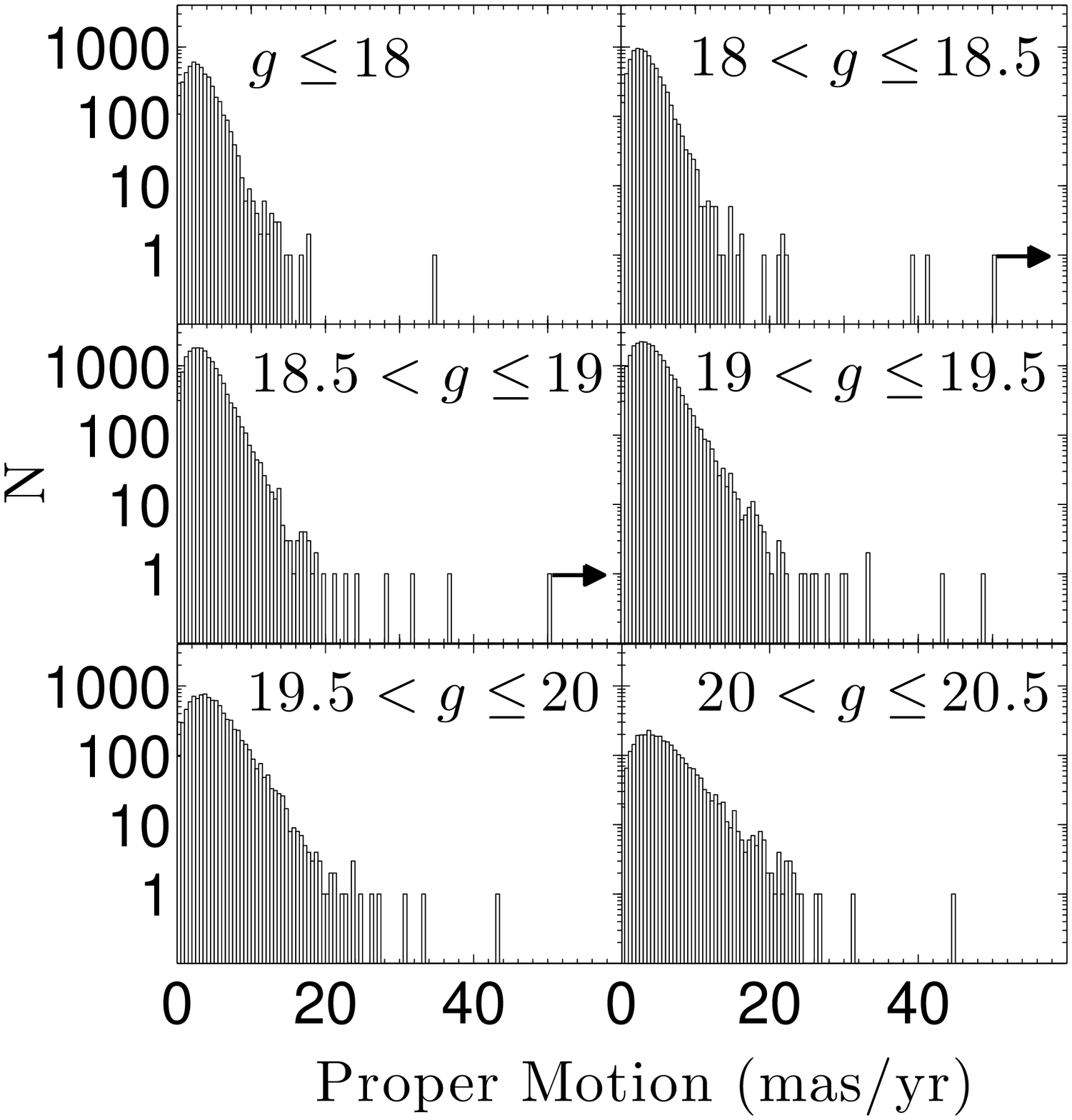}
\end{center} \figcaption{The PM distribution for S-Richards (upper left),
S-Bovy (upper right), S-Bond (lower left), and S-Schneider (lower right), 
in six bins of $g$ magnitude (uncorrected for extinction).
The arrows indicate that the last
bin contains all the ${\rm PM}\geq50$ \mpy~objects.
\label{fig:sample}}
\end{figure}

The total PM (${\rm PM}=\sqrt[]{{\tt pml}^2+{\tt pmb}^2}$, 
where {\tt pml} and {\tt pmb} are the
longitudinal and latitudinal components of PM in the
Munn et al. catalog, where {\tt pml} contains the factor of $\cos{b}$) distributions at 6 $g$-band
bins for the four samples cuter by our strict clean PM criteria are shown in Figure~\ref{fig:sample}, where the
magnitude is the $psf$ magnitude {\it without} extinction correction. The
statistical information for all the five samples is listed in Table~\ref{tab:sample}, including the number of objects with PM$\geq10$ \mpy (the commonly assumed 3$\sigma$ uncertainty), and PM$\geq30$ \mpy (defined as the ``outliers'', see below). We note that comparing with S-Schneider, S-Schneider-W is about $25\%$ larger in size, but contains over an order of magnitude more objects with PM$\geq30$ \mpy. These objects, which lie in the tail of the distribution, are almost certainly due to either contamination or failed PM measurement, as discussed above. Given the high quality of the eyeball-inspected Schneider catalog, very likely it is the latter which makes the most contribution. Again, in order to excluding these failed PM measurements as likely as possible, a strict set of clean PM conditions as ours should be applied instead of the original recommended one in \citet{mun04}.

Among the four samples resulting from our good PM criteria, in
general S-Richards (slightly better) and S-Schneider are the two cleanest, with fewer than $0.03\%$ of the objects having PM $>30$ \mpy. That is as
expected, since the spectra in the S-Schneider sample have been visually 
inspected and verified to have quasar-like spectra,
and the high {\tt good$^\prime$} index objects in S-Richards have
been cross matched with a lot of quasar-related information. In fact,
in the magnitude range covered in this study, the samples overlap almost
completely. On the other hand,
S-Bovy performs as well as the previous two samples except at the very high
PM end, while S-Bond has a substantially larger fraction of contamination 
by spurious high proper motion values.
This is due to the fact that its
selection only relies on the catalog spectroscopic redshift, which
occasionally produces a false measurement by the spectroscopic pipeline, while the
S-Schneider sample spectra were visually examined. The size of the four samples cutted by our strict good PM conditions varies
from 50,375 (S-Richards) to 66,658 (S-Schneider), while S-Bovy has too few
objects at the bright end for statistical use. For our purpose of studying
the PM error distribution, we choose S-Schneider as the main sample for
analysis, and also study S-Richards for comparison.

\begin{deluxetable}{cccccc}
\tabletypesize{\scriptsize}
\tablewidth{0pc}
\tablecaption{Sample statistics}
\tablehead{
\colhead{Number of objects in each category} &
\colhead{S-Richards} &
\colhead{S-Bovy} &
\colhead{S-Bond} &
\colhead{S-Schneider} &
\colhead{S-Schneider-W}
}

\startdata
$g\leq18.0$ & 3962 & 89 & 3893 & 4833 & 5131 \\
$18.0<g\leq18.5$ & 6371 & 2994 & 7193 & 8061 & 8504 \\
$18.5<g\leq19.0$ & 13397 & 10785 & 16076 & 17465 & 18794 \\
$19.0<g\leq19.5$ & 17862 & 17252 & 22247 & 24012 & 27549 \\
$19.5<g\leq20.0$ & 6711 & 16741 & 7846 & 9221 & 13459 \\
$20.0<g\leq20.5$ & 2072 & 7604 & 1434 & 3066 & 10075 \\
Total & 50375 & 55465 & 58689 & 66658 & 83512 \\
PM$\geq10$ \mpy & 1535	& 1900 & 2623 & 1980 & 4180 \\
PM$\geq30$ \mpy & 12 & 49 & 296 & 17 & 227 \\
PM$\geq50$ \mpy & 1 & 24 & 214 & 2 & 70 \\
\enddata

\label{tab:sample}
\end{deluxetable}



\section{The distribution of quasar PM}\label{sec:fitting}

In this Section, we fit the total PM distribution in each $g$ magnitude bin
of our quasar samples by analytic expressions, and investigate the
dependence of the fitting result on magnitude. We use the magnitude
uncorrected for extinction, since the errors should depend 
only on the apparent brightness of the source.

Equation~\ref{eq:f6} gives the core + wing
function that we use to fit the quasar PM
distributions:

\begin{equation}
f(p|A,\sigma,B,\alpha,\beta,c)=A\frac{p}{\sigma^2}e^{-\frac{p^2}{2\sigma^2}}+Bp^\alpha
e^{-(\frac{p}{c\sigma})^\beta} \label{eq:f6}
\end{equation}

We here and in what follows assume that the proper motion errors are isotropic;
we will return to this assumption below. 
We use a 2D Gaussian function to model the central part of the
distribution, where $p$ is the proper motion, $A$ is the amplitude of the
Gaussian core, and $\sigma$ is the width. For the wing part of the
distribution, we tried various fitting functions, eventually finding
that the second term in Equation~\ref{eq:f6} provides an
adequate fit.
$B$ is the amplitude,
$\alpha$ and $\beta$ are two indices, and $c$ is a constant. With values
of both
indices close to 1, this wing function decays exponentially at large PM,
as suggested by Figure~\ref{fig:sample}, and is well-behaved at
small PM with little effect on the gaussian core. For
all magnitude bins, we use PM in the range of $0-30$ \mpy~for fitting.
Extrapolation to larger proper motion values is well-behaved, though 
even in the cleanest samples the few objects at larger values have flattish
distributions and are almost
certainly either contaminants or failed measurements. Thus
we can say little about the distribution beyond 30 \mpy, except to
quote maximum probabilities for whatever causes a measurement error this 
large or larger to occur under the assumption that all the outliers
are due to measurement error, which may well be true for S-Schneider.
The fractions of these outliers in each $g$ bin are listed in Table~\ref{tab:fitting}.

\begin{figure}[tb]
\begin{center}
\epsscale{1.0} \plotone{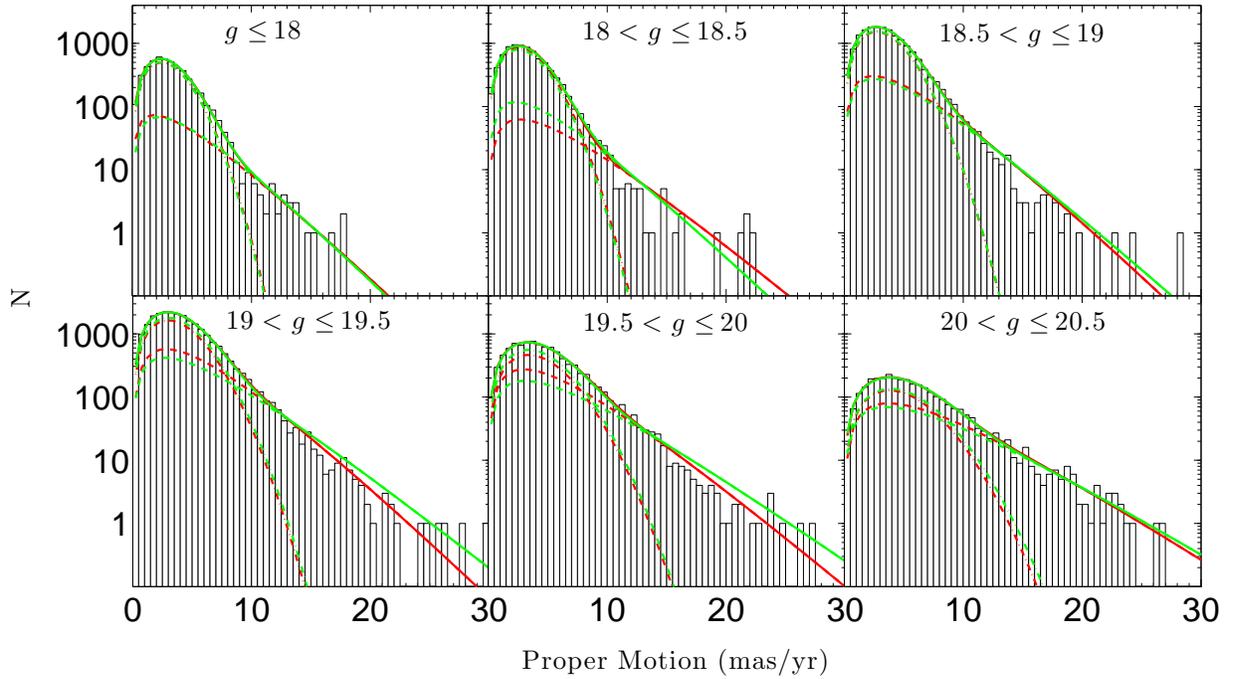}
\end{center} \figcaption{The fitting results for S-Schneider in six
bins of $g$ magnitude (uncorrected for extinction). The
red curves show the full six-parameter function
(Equation~\ref{eq:f6}), while the green curves show
the two-parameter approximation ($\alpha=1.0, \beta=1.0,
c=0.9$,  Equation~\ref{eq:f6},  and normalized $b=0.035$). 
The dash-dotted lines
are the core of the distribution (first term in 
Equation~\ref{eq:f6} and \ref{eq:f2}), the dashed lines are the wing of the distribution
(second term of Equation~\ref{eq:f6} and \ref{eq:f2}), and the solid lines are the
combination of the two.
\label{fig:fitting}}
\end{figure}

We minimize the total $\chi^2$ to fit the PM distribution at each bin to
find $A,\sigma, B, \alpha, \beta$, and $c$. We then normalize the distribution
function to get the probability function by replacing $A$ and $B$ with $a$ and
$b$ to make the integral unity ($\int_0^\infty
f(p|a,\sigma,b,\alpha,\beta,c)dp=1$). We plot the fitting results in
Figure~\ref{fig:fitting} for S-Schneider (red curves), and list the value of
the parameters and the $\chi^2$ statistics in Table~\ref{tab:fitting} for
S-Schneider and S-Richards (first two sets of rows). The fits are
generally good, with normalized $\chi^2$ in the range of 0.66-1.5 for
both samples. The fitted parameters for the two samples are similar to each
other, which indicates the robustness of the fitting function. (But
remember that the two samples are not by any means independent.)

\begin{deluxetable}{cccccccccccc}
\tabletypesize{\scriptsize}
\tablewidth{0pc}
\tablecaption{Fitting parameters and statistics}
\tablehead{
\colhead{Sample} &
\colhead{$g$ range} &
\colhead{$g_{ave}$\tablenotemark{a}} &
\colhead{$f_o$\tablenotemark{b}} &
\colhead{$a$\tablenotemark{c}} &
\colhead{$\sigma$} &
\colhead{$b$\tablenotemark{d}} &
\colhead{$\alpha$} &
\colhead{$\beta$} &
\colhead{$c$} &
\colhead{$\chi^2$} &
\colhead{Normalized $\chi^2$}
}

\startdata
S-Richards\tablenotemark{e} & $g\leq18.0$ & 17.52 & 0.000\% & 0.92 & 2.48 & 0.020 & 0.51 & 0.88 & 0.97 & 43.2 & 0.80 \\
 & $18.0<g\leq18.5$ & 18.28 & 0.031\% & 0.88 & 2.53 & 0.016 & 1.03 & 1.03 & 0.93 & 59.7 & 1.11 \\
 & $18.5<g\leq19.0$ & 18.78 & 0.015\% & 0.78 & 2.72 & 0.038 & 0.89 & 1.03 & 1.01 & 67.5 & 1.25 \\
 & $19.0<g\leq19.5$ & 19.24 & 0.022\% & 0.71 & 2.99 & 0.036 & 1.10 & 1.06 & 1.06 & 80.9 & 1.50 \\
 & $19.5<g\leq20.0$ & 19.70 & 0.045\% & 0.57 & 3.50 & 0.054 & 1.01 & 1.00 & 1.25 & 45.9 & 0.85 \\
 & $20.0<g\leq20.5$ & 20.19 & 0.048\% & 0.46 & 3.76 & 0.039 & 1.11 & 1.01 & 1.10 & 40.8 & 0.76 \\
 &  &  &   &  &  &  &  &  &  &  &  \\
S-Schneider\tablenotemark{e} & $g\leq18.0$ & 17.53 & 0.021\% & 0.83 & 2.44 & 0.041 & 0.77 & 1.01 & 0.98 & 39.4 & 0.73 \\
 & $18.0<g\leq18.5$ & 18.28 & 0.037\% & 0.89 & 2.53 & 0.016 & 1.00 & 1.01 & 1.06 & 68.3 & 1.27 \\
 & $18.5<g\leq19.0$ & 18.78 & 0.017\% & 0.77 & 2.70 & 0.037 & 0.93 & 1.05 & 1.00 & 81.1 & 1.50 \\
 & $19.0<g\leq19.5$ & 19.24 & 0.021\% & 0.67 & 2.98 & 0.044 & 1.14 & 1.04 & 0.86 & 74.6 & 1.38 \\
 & $19.5<g\leq20.0$ & 19.70 & 0.033\% & 0.56 & 3.38 & 0.053 & 1.14 & 1.01 & 0.78 & 62.3 & 1.15 \\
 & $20.0<g\leq20.5$ & 20.20 & 0.065\% & 0.52 & 3.80 & 0.036 & 1.14 & 1.00 & 0.86 & 35.4 & 0.66 \\
 &  &  &   &  &  &  &  &  &  &  &  \\
S-Schneider\tablenotemark{f} & $g\leq18.0$ & 17.53 & 0.021\% & 0.83 & 2.42 & 0.035 & 1.00 & 1.00 & 0.90 & 40.9 & 0.71 \\
 & $18.0<g\leq18.5$ & 18.28 & 0.037\% & 0.82 & 2.51 & 0.035 & 1.00 & 1.00 & 0.90 & 74.9 & 1.29 \\
 & $18.5<g\leq19.0$ & 18.78 & 0.017\% & 0.79 & 2.70 & 0.035 & 1.00 & 1.00 & 0.90 & 81.7 & 1.41 \\
 & $19.0<g\leq19.5$ & 19.24 & 0.021\% & 0.74 & 3.01 & 0.035 & 1.00 & 1.00 & 0.90 & 93.4 & 1.61 \\
 & $19.5<g\leq20.0$ & 19.70 & 0.033\% & 0.68 & 3.38 & 0.035 & 1.00 & 1.00 & 0.90 & 79.4 & 1.37 \\
 & $20.0<g\leq20.5$ & 20.20 & 0.065\% & 0.57 & 3.91 & 0.035 & 1.00 & 1.00 & 0.90 & 38.6 & 0.67 \\
 &  &  &   &  &  &  &  &  &  &  &  \\
S-Schneider-W\tablenotemark{e} & $g\leq18.0$ & 17.53 & 0.12\% & 0.76 & 2.44 & 0.056 & 0.99 & 1.05 & 0.90 & 41.5 & 0.77 \\ & $18.0<g\leq18.5$ & 18.28 & 0.15\% & 0.85 & 2.53 & 0.022 & 1.05 & 1.04 & 1.03 & 55.2 & 1.02 \\ & $18.5<g\leq19.0$ & 18.78 & 0.13\% & 0.78 & 2.75 & 0.038 & 0.89 & 1.01 & 0.94 & 84.9 & 1.57 \\ & $19.0<g\leq19.5$ & 19.24 & 0.18\% & 0.75 & 3.08 & 0.034 & 0.97 & 1.00 & 0.90 & 134.3 & 2.49 \\ & $19.5<g\leq20.0$ & 19.72 & 0.35\% & 0.67 & 3.49 & 0.031 & 0.99 & 0.99 & 0.93 & 110.7 & 2.05 \\ & $20.0<g\leq20.5$ & 20.24 & 0.87\% & 0.46 & 4.18 & 0.033 & 1.12 & 1.00 & 0.88 & 71.6 & 1.33 \\
\enddata

\tablenotetext{a}{Average $g$ magnitude of objects in each bin.}
\tablenotetext{b}{Fraction of the outliers (objects with $pm\geq30$ \mpy).}
\tablenotetext{c}{Normalized $A$ in Equation~\ref{eq:f6}.}
\tablenotetext{d}{Normalized $B$ in Equation~\ref{eq:f6}.}
\tablenotetext{e}{Fitted by Equation~\ref{eq:f6}.}
\tablenotetext{f}{Fitted by Equation~\ref{eq:f2} with one more free parameter on the overall scale.}
\label{tab:fitting}
\end{deluxetable}

\begin{figure}[tb]
\begin{center}
\epsscale{1.0} \plotone{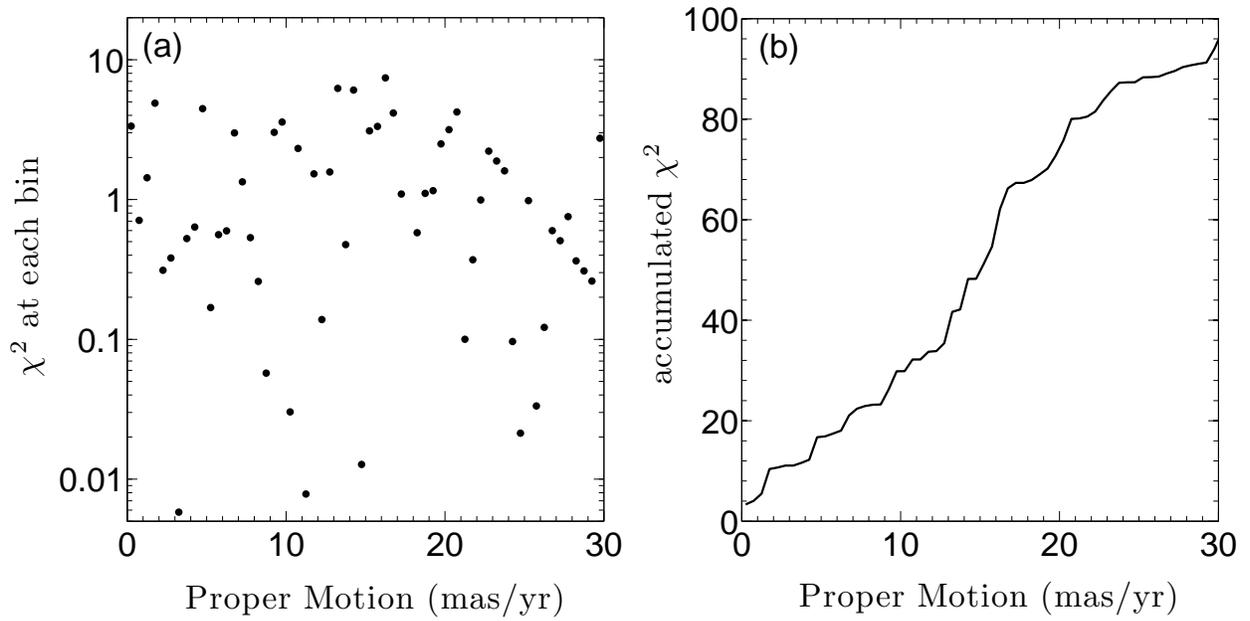}
\end{center} \figcaption{(a): The $\chi^2$ at each PM bin for the freedom
reduced fitting for the $18<g\leq18.5$ bin in S-Schneider. (b): the
accumulated $\chi^2$ in this case (the degree of freedom is 58).
\label{fig:chi2}}
\end{figure}

It is clear that several of the parameters do not change very much
with magnitude, so we also conducted experiments in which we
freeze the values of some of these, and refit the distributions with
this reduced freedom.
The third set of rows
in Table~\ref{tab:fitting} shows the best result from this exercise, where we
fix $\alpha=1.0,~\beta=1.0,~c=0.9$, define a normalized tail amplitude
$b=0.035$ (and a corresponding normalized gaussian amplitude
$a=1-b(c\sigma)^2$). This makes the normalized probability function a
one-parameter function of $\sigma$ alone; ($\int_0^\infty f(p|\sigma)dp=1$).
Figure~\ref{fig:fitting} shows these results (green
curves), where it is clear that the difference between the full-freedom and 
the reduced-freedom
fitting is small in all the magnitude bins. The total $\chi^2$ only
moderately increases with the new fitting function (the normalized $\chi^2$
even drops in some cases due to an increase in the degree of freedom), and the largest increase
occurs at the $19.0<g\leq19.5$ magnitude bin (which also has the largest
normalized $\chi^2$ (1.61) and the largest population, so is most
sensitive to small inadequacies in the fitting function). 
We show the detailed $\chi^2$ map for this case in
Figure~\ref{fig:chi2}. The $\chi^2$ at each PM bin are uniformly scattered
around 1, and the accumulated $\chi^2$ behaves well.
The normalized one-parameter fitting function is:

\begin{equation}
f(p|\sigma)=(1-0.035(0.9\sigma)^2)\frac{p}{\sigma^2}e^{-\frac{p^2}{2\sigma^2}}+
0.035pe^{-(\frac{p}{0.9\sigma})}
\label{eq:f2}
\end{equation}

Due to the limited quasar sample size we can only fit its PM distribution
and extract the fitted parameters at six $g$ magnitude bins (the average
magnitude of each $g$ bin for both S-Schneider and S-Richards are listed in
Table~\ref{tab:fitting}). To get the distribution function at some arbitrary
$g$, we fit a quadratic function to the $\sigma$ in the fitting using the
one-parameter fits
(Equation~\ref{eq:f2}, the only free parameter in the normalized
function) as a function of $g$:

\begin{equation}
\sigma=0.2293(g-19)^2 + 0.6205(g-19) + 2.836
\label{eq:sigma-g}
\end{equation}

\begin{figure}[tb]
\begin{center}
\epsscale{0.49} \plotone{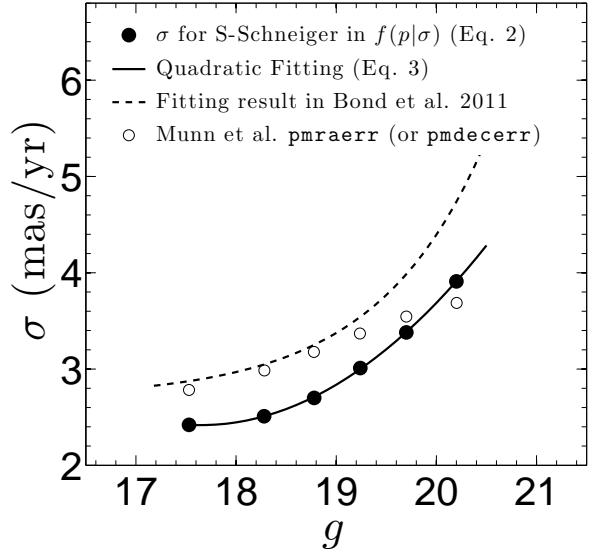}
\end{center}
\figcaption{The value of $\sigma$ in the approximate distribution
(Equation~\ref{eq:f2}) of the quasar PM in S-Schneider as a
function of average $g$ magnitude (solid point, the third set of rows in
Table~\ref{tab:fitting}), and the corresponding fitting function (solid line, Equation~\ref{eq:sigma-g}). The dashed line shows the $\sigma$ fitting result in \citet{bon10} (their Equation 1. We convert the $r$ magnitude into $g$ magnitude by $g-r=0.18$, the average value for S-Schneider.). The open circles are the average catalog-provided PM error estimate in Munn et al. catalog ({\tt pmraerr} or {\tt pmdecerr}) for each $g$ bin.
\label{fig:sigma-g}}
\end{figure}

Figure~\ref{fig:sigma-g} shows the fitting results, which are very good. \citet{bon10} studied the PM error distribution using their quasar sample. They fitted a Gaussian profile to the error distribution in the entire magnitude range, and obtained a fitting function for $\sigma$, which is shown here as well for comparison (They fitted $\sigma$ as a function of $r$-band magnitude. To make a direct comparison we convert $r$ into $g$ via $g-r=0.18$, which is the average value for S-Schneider, see below.). In general, their fitted $\sigma$ is larger (by $\sim0.5$ \mpy) than ours in the entire magnitude range, while the two curves share very similar shapes. This is fully consistent with the fact that \citet{bon10} used a less clean quasar sample than S-Schneider and didn't separate the tail of the distribution from the Gaussian core, both resulting in a larger Gaussian width.
Our fitting formula, of course, applies only to a finite range in
g magnitude, since we have only a finite range over which to determine 
it. We recommend using
it in the range of $17.5\leq g\leq20.5$ (the range we plot in the figure).
For $g<17.5$ we recommend using a constant $\sigma=2.42$, because for bright
objects the proper-motion measurement error approaches the instrument induced
error limit and does not depend on the brightness of the object. For
$g>20.5$, the fitting function is ill-constrained; we have too few
quasars to calibrate it, and the core parameters are not well-determined.

\begin{figure}[tb]
\begin{center}
\epsscale{0.49} \plotone{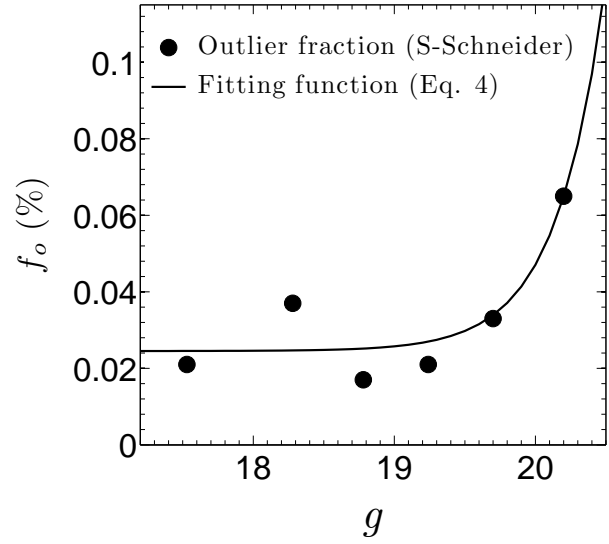}
\end{center}
\figcaption{The fitted outlier fraction $f_o$ (Table~\ref{tab:fitting})
in S-Schneider as a function of $g$ (Equation~\ref{eq:fo-g}).  \label{fig:fo-g}}
\end{figure}

Finally, we fit the fraction of the outliers ($f_0$ in 
Table~\ref{tab:fitting}) in S-Schneider as a function of $g$ magnitude. 
Given the small number of outliers (17), this fitting can only indicate
the general trend of the $f_o-g$ relation. We approximate the fitting
function by:
\begin{equation}
f_o=0.00122(10^{1.27(g-19)}+20.1) \%
\label{eq:fo-g}
\end{equation}
(Note $f_o$ is in unit of $\%$.), and the fitting is shown in Figure \ref{fig:fo-g}.
As is the case with the fitted $\sigma-g$
relation, we recommend confining the use of this relation to
$g\lesssim20.5$ since we have insufficient numbers of objects at
fainter magnitudes.  The extension to the brighter magnitudes
can probably be trusted, however. 

\begin{figure}[tb]
\begin{center}
\epsscale{1.0} \plotone{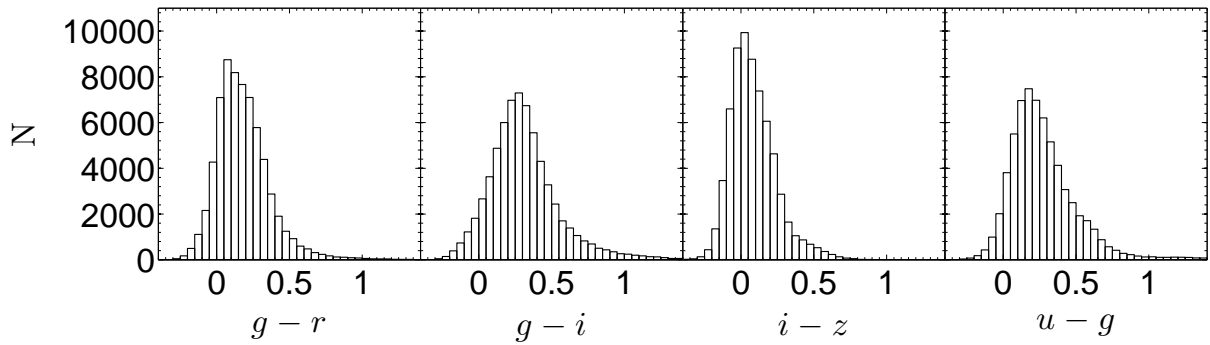}
\end{center} \figcaption{Histograms of colors (no extinction correction) for
S-Schneider.
\label{fig:color}}
\end{figure}

\begin{figure}[tb]
\begin{center}
\epsscale{0.5} \plotone{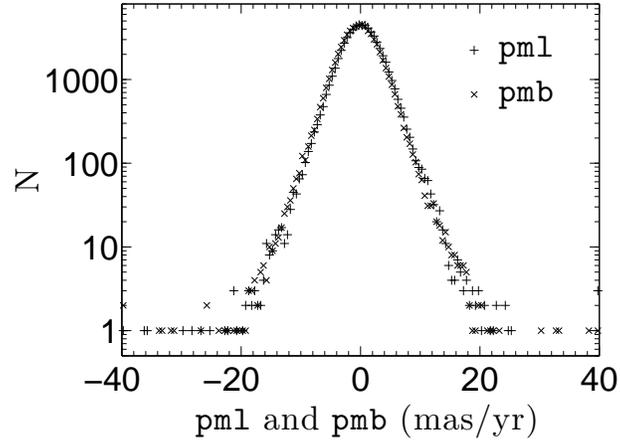}
\end{center}
\figcaption{The distribution of PM in each component ({\tt pml} and {\tt pmb}) in
the Munn et al. catalog for S-Schneider.  \label{fig:pmlpmb-histo}}
\end{figure}

Before we move to the next section, there are several general issues about the PM
error distribution that we would like to discuss.
\begin{enumerate}

\item Since most previous studies which used the SDSS+USNO-B PM catalog employed the original clean PM criteria in \citet{mun04} to select objects with good PM measurements, here we explore the effect of this weaker PM cut on the error distribution by fitting S-Schneider-W using Equation~\ref{eq:f6}. The result is shown in Table \ref{tab:fitting} (bottom set of rows). Comparing with the fitting results of S-Schneider (the second set of rows in the same table), the original clean PM conditions results in a slightly larger $\sigma$ (up to $\sim10\%$), and a generally worse $\chi^2$ statistics (both are more significant at the faint end). The biggest disadvantage of the original clean PM criteria is still, as we discussed above, that it introduces a much larger $f_o$. Studies of stellar kinematics which are sensitive to the outlier fractions, such as searching for extremely high velocity stars in the Galaxy, will be severely affected if this weaker set of clean PM criteria is applied instead of ours.

\item In principle, PM errors could depend on color as well as magnitude;
\citet{bon10} found that the
systematic errors in PM have a small color dependence, but did not
find a corresponding dependence for the random errors. Basically, the
color range of quasars is too small to investigate this
possibility. Several color distributions for S-Schneider are shown in
Figure \ref{fig:color}. Stars, on the other hand, have a much wider range
of colors and their measured proper motions may be subject to 
color-related errors which cannot be investigated with quasar samples.
If this is the case, the error functions derived here are more applicable 
to samples of blue stars, such as main-sequence-turnoff samples. 
The POSS positions which enter the proper motion calculation are
inverse-variance weighted combinations of data from the O, J, E, F, and
N plates and thus span a very large wavelength range, most of which is
to the red of the effective wavelength of the SDSS $g$ band, but it is
{\it not clear} exactly what the effective wavelength is.  But we probably
somewhat overestimate the errors for red objects, because they are
brighter in most of the photographic bands than typical quasars with
the same $g$ magnitude, {\it i.e.} our result is a conservative limit for
red objects (which is reason that we choose to investigate the dependence
of the PM error distribution on $g$ magnitude instead of on redder bands). The mean $g-r$ color in the figure is 0.18,
which corresponds to middle-late F stars.  The sample we are
investigating which prompted this study is a SEGUE II sample of halo
turnoff stars, for which the color match with the quasars is (entirely
fortuitously) excellent. 

\item There are additional possible contributions to the PM errors,  
such the way of measuring the position of the objects on the sky and observing conditions.
The USNO-B positions are originally in the coordinate system of the USNO
plates and then later transferred to the celestial coordinates (right
ascension and declination), while the SDSS positions are initially measured 
in the CCD coordinate system, which later are transferred to the
survey longitude and latitude which define the photometric scans
of the sky, and are then calibrated with respect to the ra-dec measurements
from UCAC. The derivation of proper motions from these two sets of 
measurements has the probable effect of making the errors in the
proper motions more nearly isotropic (See point 3 below).
\citet{bon10}
looked at the dependence of median and rms errors for the longitudinal and
latitudinal PM components on position on the sky, and concluded that the
variation is relatively small (with 
the median variation being much smaller than $\sigma$).
On the other hand, whether the PM error distribution (especially the wing
component)
depends systematically on position on the sky is another question, 
which again we are
not able to address due to the limited size of the quasar samples. Note,
however, that our investigation deals well with the aggregate survey,
so questions like the number of outliers expected with samples
large enough to sample the sky in a manner comparable to the quasars
should be well answered.

\item We investigate the error distribution for the total PM and not
for the individual components of the PM, since the errors
are likely to be isotropic and, for many purposes,
it is the total PM which matters.
\citet{bon10} concluded that the correlation between
the errors in the two components is negligible compared to the total random
and systematic errors. In addition, we find that the error
distributions in the two components are very similar, as shown in
Figure~\ref{fig:pmlpmb-histo}. 
For S-Schneider,
the median is $0.10$ for {\tt pml} and $-0.17$ for {\tt pmb}, both significantly
smaller than the Gaussian width $\sigma$, and the standard deviation is 3.40
for {\tt pml} and 3.43 for {\tt pmb}. This inferred near-isotropy of the PM errors, 
however, should be viewed with some caution, as the l and b components
are related in a very complex and variable way to components either
along and perpendicular to the scan directions in SDSS, altitude and
azimuth, or right ascension and declination, for which in any of those
cases there might be factors contributing to anisotropy.

\item In addition to providing the PM measurement, \citet{mun04} also provided
a PM error estimate ({\tt pmraerr} and {\tt pmdecerr}, which represent the expected standard deviations of the PM measurement around the true value in each direction, but the two components are always assumed to be the same in the catalog). Since most previous investigations which used the SDSS stellar kinematics to study the galactic structure employed the catalog-provided {\tt pmraerr} and {\tt pmdecerr} to estimate the uncertainty of the PM measurement, it is worth calibrating the performance of this error estimate using the true PM error distribution which we get from our quasar samples.
We calculate the total catalog-provided PM error estimate (${\rm PM}_{\rm
error}=\sqrt{{\tt pmraerr}^2+{\tt pmdecerr}^2}$) for S-Schneider, and show its statistical distribution in Figure~\ref{fig:munn-pmerror}. The distribution is rather narrow, effectively ranging from $3-6$ \mpy. In addition, we provide the average {\tt pmraerr} (or {\tt pmdecerr}) for each $g$ bin, and over plot it in Figure \ref{fig:sigma-g} (from small to large $g_{avg}$, the values are 2.78, 2.99, 3.18, 3.37, 3.54, and 3.69 \mpy). We note here although we use a 2D Gaussian function to model the core part, the $\sigma$ in the PM error pdf is still a 1D Gaussian $\sigma$ ({\it i.e.} in one direction, assuming isotropy for the distribution), so it is {\tt pmraerr} (or {\tt pmdecerr}) which should be compared with the fitted $\sigma$, not ${\rm PM}_{\rm error}$. The comparison shows that the catalog-provided
error estimate is in good agreement with our fitted $\sigma$ (within $20\%$), with the former on average being $\sim12\%$ higher than the latter (the agreement is better at the faint end). The major drawback of the catalog-provided error estimate is that it does not address the non-Gaussian tail of the error distribution.
In the further, we recommend that analysis
using the Munn et al. PM, especially the ones for which the tale of the PM error distribution is important, should use our results to estimate the error instead of using the catalog-provided errors.
\end{enumerate}



\section{The Significance of the measured proper
motions}\label{sec:significance}

\begin{figure}[tb]
\begin{center}
\epsscale{0.49} \plotone{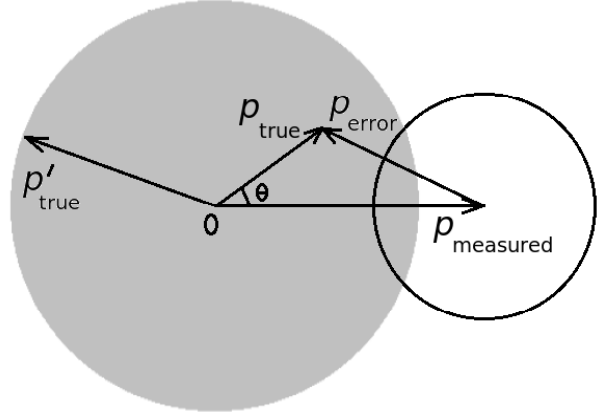}
\end{center} \figcaption{Schematic plot which shows how we calculate the
probability $F(p_{\rm true}\leq p_{\rm true}^\prime)$.
\label{fig:schematic}}
\end{figure}

With the normalized PM distribution function $f(p)$ for our quasar samples,
we can calculate 
the PM error probability function ($f(p_{\rm error})=f(p)$) of the
measured PM ($p_{\rm measured}$) for an object (a star) with intrinsic
non-zero PM. Specifically, $f(p_{\rm error})dp_{\rm error}$ is the probability that the
true proper-motion $\vec{p}_{\rm true}=\vec{p}_{\rm measured}+\vec{p}_{\rm
error}$ falls in an annulus centered on $\vec{p}_{\rm measured}$ with radius
$p_{\rm error}$ and $p_{\rm error}+dp_{\rm error}$. (see
Figure~\ref{fig:schematic}). Based on this, we could calculate the probability
of an object with $\vec{p}_{\rm measured}$ having an (unknown) true PM
$p_{\rm true}$ larger than some certain value $p_{\rm true}^\prime$. As shown
in Figure~\ref{fig:schematic}, the total probability of $p_{\rm true}$ being
in the shadowed area ($p_{\rm true}\leq p_{\rm true}^\prime$) is:

\begin{equation}
F(p_{\rm true}\leq p_{\rm true}^\prime)=\int_0^{p_{\rm
true}^\prime}\frac{f(p_{\rm error})}{2\pi p_{\rm error}}pdpd\theta
\label{eq:probability}
\end{equation}

While $\vec{p}_{\rm error}=\vec{p}_{\rm true}-\vec{p}_{\rm measured}$ so
$p_{\rm error}=\sqrt[]{p_{\rm measured}^2+p_{\rm true}^2-2p_{\rm
measured}p_{\rm true}\cos{\theta}}$. The above integral turns into:

\begin{equation}
F(p_{\rm true}\leq p_{\rm true}^\prime)=\int_0^{p_{\rm
true}^\prime}\frac{f(\sqrt[]{p_{\rm measured}^2+p^2-2p_{\rm
measured}p\cos{\theta}})}{2\pi \sqrt[]{p_{\rm measured}^2+p^2-2p_{\rm
measured}p\cos{\theta}}}pdpd\theta \label{eq:probability2}
\end{equation}

\begin{figure}[tb]
\begin{center}
\epsscale{1.0} \plotone{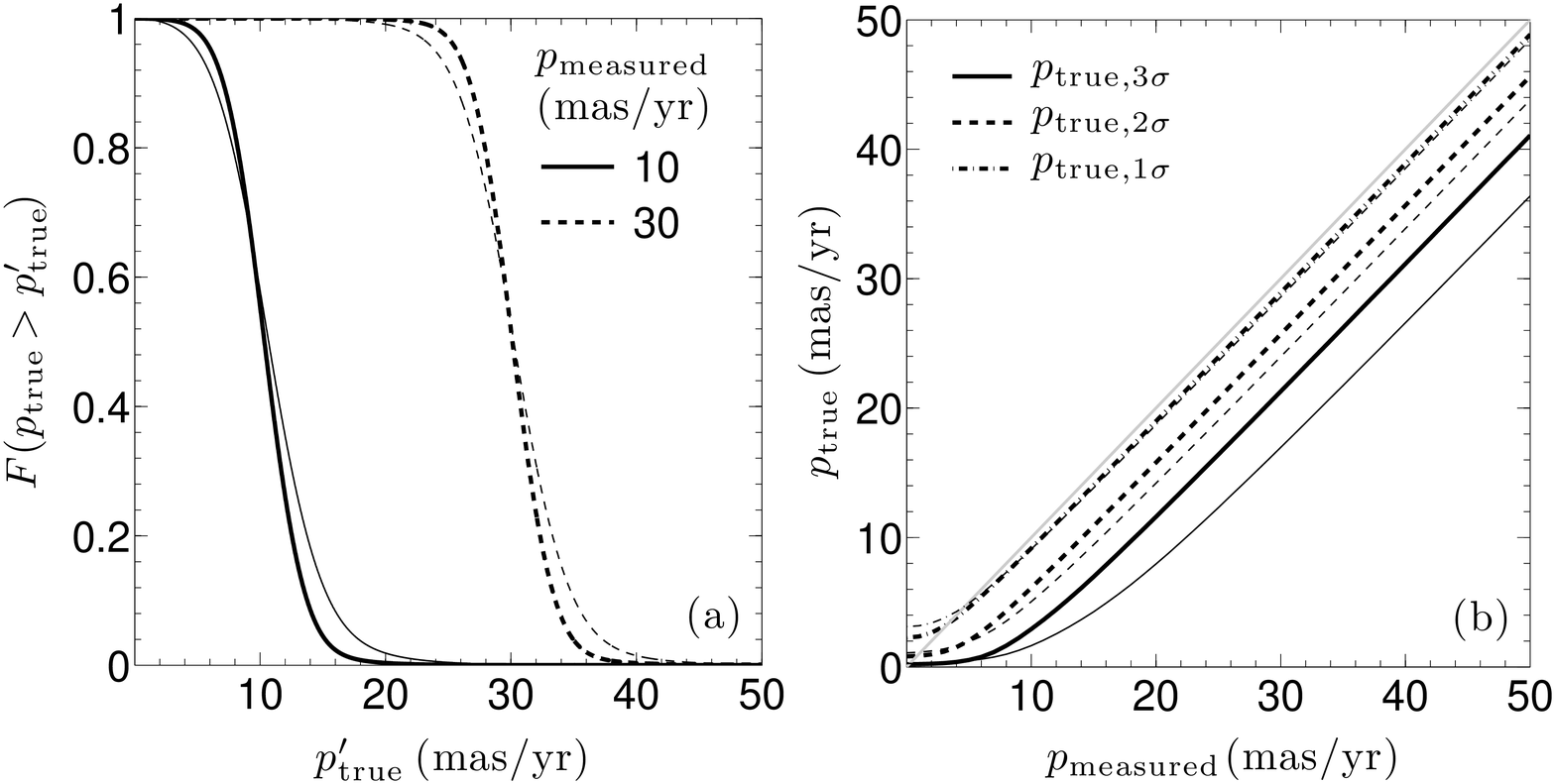}
\end{center} \figcaption{(a): The calculated probability $F(p_{\rm
true}>p_{\rm true}^\prime)$ (Section~\ref{sec:significance}) as a function of
$p_{\rm true}^\prime$ for two measured PM at 10 and 30 \mpy. (b): Three
confidence levels for the true PM (Section~\ref{sec:significance}) as a
function of measured PM. In both plots thick lines are for $g=18.28$, and
thin lines are for $g=19.70$. Here 1, 2, and 3 $\sigma$ refer to probabilities
of 68\%, 95\%, and 99.7\% that the true PM exceeds the plotted value.
\label{fig:probability}}
\end{figure}

Thus $F(p_{\rm true}>p_{\rm true}^\prime)=1-F(p_{\rm true}\leq p_{\rm
true}^\prime)$ is the probability that $p_{\rm true}$ falls outside the
shadowed area, {\it i.e.} the object in consideration has a true PM larger
than the given threshold $p_{\rm true}^\prime$. Panel $(a)$ in
Figure~\ref{fig:probability} shows the calculation of $F(p_{\rm true}>p_{\rm
true}^\prime)$ for two measured PM at two $g$ magnitudes ($18.28$ and
$19.70$). Based on this, we define a series of confidence intervals 
($p_{\rm true,1\sigma}$, $p_{\rm true,2\sigma}$, and $p_{\rm true,3\sigma}$)
to be the $p_{\rm true}^\prime$ corresponding to $F(p_{\rm true}>p_{\rm
true}^\prime)=68\%$, $95\%$, and $99.7\%$. Panel $(b)$ in
Figure~\ref{fig:probability} shows these confidence
intervals.

Here we use an example to illustrate the effect of the non-Gaussian tail in the PM error distribution in determining the confident velocity of a star with some measured PM. For a star with measured PM of 30 \mpy~(corresponding to a transverse velocity of 423 km s$^{-1}$ at a helio-centric distance of 3 kpc) and $\sigma$ of 4 \mpy, if the PM error distribution only contains the Gaussian core ({\it e.g.} $f(p|\sigma)=(p/\sigma^2)e^{-p^2/(2\sigma^2)}$), the $2\sigma$ confident PM would be 23.7 \mpy (334 km s$^{-1}$) and the $3\sigma$ PM$=19.3$ (272 km s$^{-1}$). On the other hand, when including the non-Gaussian tail, the 2$\sigma$ and 3$\sigma$ confident PM decrease to 22.6 and 13.8 \mpy (319 and 195 km s$^{-1}$. When combining with a model of the Galactic potential and a radial velocity, these differences in confident velocity may flip the conclusion that whether this star is bonded to the Galaxy or not. In general, the effect of the non-Gaussian tail is more prominent at the faint end, where the Gaussian width is larger and the weight of the non-gaussian tail is bigger.

Lastly, we note that in this section all the calculations are done with the
one-parameter error probability function $f(p_{\rm error})=f(p|\sigma(g))$
(Equation~\ref{eq:f2} and Table~\ref{tab:fitting}), but it is a trivial
exercise to replace $f(p|\sigma)$ with $f(p|a,\sigma,b,\alpha,\beta,c)$
(Table~\ref{tab:fitting}) if more accurate results are desired. However, in
doing so one needs to interpolate the parameters at the six discrete $g$
magnitudes to get the result at some arbitrary $g$.



\section{Summary and discussion}\label{sec:summary}

We have investigated the proper-motion measurement errors in the SDSS+USNO-B
proper-motion catalog \citep{mun04} by analyzing the proper-motion
distributions of several recent SDSS quasar samples
\citep{ric09,bon10,sch10,bov11}. The sample defined by
\citet{sch10}was determined to be the cleanest sample
and has the largest sample size. We bin the data into six
$g$ magnitude (not extinction corrected) bins, and fit analytic functions for
the PM distribution in each bin. We find that, while a six-parameter
fitting formula (Equation~\ref{eq:f6}) describes the quasar PM distribution
well, a simpler (normalized) function with one free parameter
(Equation~\ref{eq:f2}) also gives reasonably good results. Based on 
this fitting function for the PM error distribution, we calculate the
probability that an object with a measured PM has a true PM $p_{\rm
true}$ larger than a given threshold $p_{\rm true}^\prime$. Cutting
the probability $F(p_{\rm true}>p_{\rm true}^\prime)$ at several confidence
levels, we calculate the ``most likely PM'' for a given measured PM.

The analysis raised several issues which we would like to stress:

\begin{enumerate}
\item Our PM error analysis can only be applied to {\it clean} PM subsamples from 
the SDSS+USNO-B catalog, which satisfies our strict clean PM criteria ({\tt match = 1}, {\tt sigRa < 525 \&\& sigDec < 525}, {\tt dist22 > 7}, and most importantly, {\tt nFit = 6}, which requires that the object was
detected on all five USNO-B plates). Experiments show that weakening this set of
conditions (specifically, the original criteria in \citet{mun04}: {\tt match = 1} and {\tt sigRa < 350 \&\& sigDec < 350}) generally results in
a broader Gaussian core width (up to $10\%$), a worse $\chi^2$ statistics in the fitting, and a significantly higher fraction of the outliers (by an order of magnitude). Studies which are sensitive to the tail of the PM error distribution and the fraction of outliers, such as searching for extremely high velocity stars in the Galaxy, will be severely affected if using this weaker set of conditions instead of ours.

\item The PM error distribution derived here applies to
$g$ magnitudes (without extinction correction) in the range
$\sim17-20.5$. Specifically, the fitting functions Equation~\ref{eq:f2} and
\ref{eq:sigma-g} are only good for $17.5\leq g\leq20.5$. Though using
a constant $\sigma$ for brighter objects is probably appropriate,
using a constant or extrapolating our results for fainter magnitudes is not
(Section~\ref{sec:fitting}).

\item While we have derived a fitting function for the entire PM error
range, we note that there is a small fraction of ``outliers'' (defined here
to be PM error $>30$ \mpy), which apparently do not belong to the 
derived error distribution.
Even for our cleanest quasar sample, which has passed spectroscopic
visual inspection \citep{sch10}, and with our very conservative PM culling
conditions, there is still a small number
of outliers, up to $\sim0.1\%$, in the faintest magnitude bands. We do
not speculate on the origin of these, but they are likely present
in any sample chosen by our criteria.

\item While we fit the PM error distribution as a function of magnitude,
the readers should bear in mind that the PM error may depend on other
parameters as well, such as color and position of the objects on they sky. The PM error
distribution in this work is best applicable to objects in the color range
similar to the quasar sample we use to get the distribution (i.e., blue stars, Fig. \ref{fig:color}),
though our result could be considered as a conservative
limit for redder objects.
While the median and rms for the PM error only weakly depend
on the position on the sky, we do not have a large
enough sample to investigate the dependence of the distribution on the position on
the sky. The distributions of the PM error components in the longitudinal and
latitudinal are very similar to each other (Fig. \ref{fig:pmlpmb-histo}), and the correlation between the two
is negligible compared to the total random and systematic errors.

\item Last, we note that the PM error estimate provided by \citet[{\tt pmraerr}
and {\tt pmdecerr}]{mun04} is in rough agreement with the $\sigma$ fitted from the quasar PM distribution (Fig. \ref{fig:sigma-g}). The agreement is within $20\%$, and better at the faint end. On the other hand, the major drawback of the catalog-provided error estimate is that it does not address the non-Gaussian tail.
We recommend that future analysis
using Munn et al. proper motions, especially the ones for which the tail of the error distribution is important, should use this work to estimate the error instead of using the catalog-provided error estimate.
\end{enumerate}

This investigation of the PM error are useful in many ways. For example, we are studying the kinematics of stars in SEGUE II, focusing on these hyper velocity ones with velocity exceeding the escape velocity of the Galactic potential. In this case, it is crucial to have an accurate PM error distribution to judge the significance of the hyper velocity for these escapers.


\section*{Acknowledgments}

We thank Deokkeun An, Steve Bickerton, Jo Bovy, Timothy Brandt, \v{Z}eljko Ivezi\'c, Craig Loomis,
Heather Morrison, Jeff Munn, Gordon Richards, Donald
Schneider, Ralph Sch\"onrich, Michael Strauss,
and an anonymous referee for useful discussions and comments.

Funding for the SDSS and SDSS-II has been provided by the Alfred P. Sloan
Foundation, the Participating Institutions, the National Science Foundation,
the U.S. Department of Energy, the National Aeronautics and Space
Administration, the Japanese Monbukagakusho, the Max Planck Society, and the
Higher Education Funding Council for England. The SDSS Web Site is
http://www.sdss.org/. The SDSS is managed by the Astrophysical Research
Consortium for the Participating Institutions. The Participating Institutions
are the American Museum of Natural History, Astrophysical Institute Potsdam,
University of Basel, University of Cambridge, Case Western Reserve University,
University of Chicago, Drexel University, Fermilab, the Institute for Advanced
Study, the Japan Participation Group, Johns Hopkins University, the Joint
Institute for Nuclear Astrophysics, the Kavli Institute for Particle
Astrophysics and Cosmology, the Korean Scientist Group, the Chinese Academy of
Sciences (LAMOST), Los Alamos National Laboratory, the Max-Planck-Institute
for Astronomy (MPIA), the Max-Planck-Institute for Astrophysics (MPA), New
Mexico State University, Ohio State University, University of Pittsburgh,
University of Portsmouth, Princeton University, the United States Naval
Observatory, and the University of Washington.


\begin{figure}[tb]
\begin{center}
\epsscale{0.45} \plotone{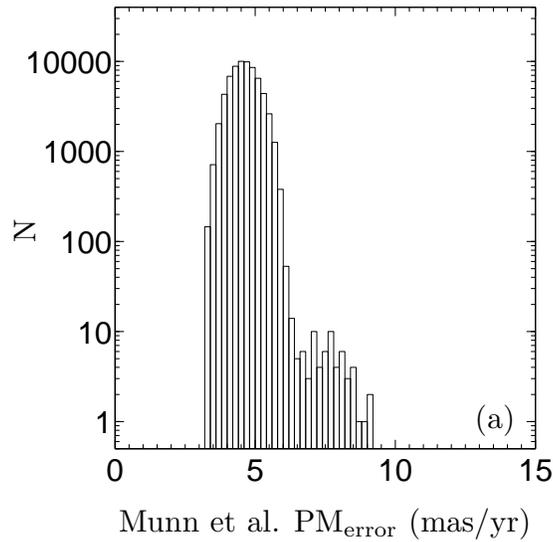}
\end{center}
\figcaption{The distribution of the PM error estimate (${\rm PM}_{\rm
error}=\sqrt{{\tt pmraerr}^2+{\tt pmdecerr}^2}$) in the Munn et al. catalog for S-Schneider.
\label{fig:munn-pmerror}}
\end{figure}

\end{document}